**Lateral force microscopy calibration using an interferometric atomic force microscope**

Joel A. Lefever, Aleksander Labuda, and Roger Proksch

*Asylum Research, an Oxford Instruments Company, Santa Barbara, CA 93117*

**Abstract**

A new method is introduced for calibrating lateral force as measured by an atomic force microscope (AFM), making use of both an interferometric detector and an optical beam detector on the same instrument. The method may be implemented automatically and performed with minimal user input. The microscope has the capability to measure the probe tip height *in situ*, which allows for a complete lateral force calibration without changing the sample or probe. Two options for lateral force measurements are described wherein the two detectors are alternately used to measure normal and lateral forces, and methods for applying the calibration protocol for both alternatives are provided. The tip height measurement is validated by direct comparison with an electron micrograph and they are generally in agreement to within 1.4 µ m. For most cantilevers tested, the complete lateral calibration method is consistent with the wedge calibration method to within the intrinsic uncertainty of the wedge method.

Lateral force microscopy (LFM) is used to measure friction forces in various materials including two-dimensional materials [1–3], ceramics [4], and self-assembled monolayers [5]. However, in many cases, the lateral signal is reported in arbitrary units due to the challenges associated with calibration. While the sensitivity of the photodetector to vertical motion of the cantilever tip may be estimated by pressing the cantilever into a hard surface, the lateral sensitivity of the quadrant photodetector cannot be easily calibrated in a similar manner because the probe tip may slide across the surface under horizontal loading. Numerous schemes [6, 7] have been identified to circumnavigate this challenge, including methods based on sliding across a sloped surface [8–10], methods based on hydrodynamics [11, 12], a method in which the tip is brought into contact with a diamagnetically levitated sample [13], and methods in which colloids or fibers are added to the test probe [14] or a reference probe [15]. However, all of these existing calibration methods are complex and feature substantial limitations, typically involving exchanging the sample of interest for a calibration sample, being restricted to use in air, or being limited to particular cantilever geometries or designs. Moreover, some methods have the potential to damage or destroy the cantilever, for example by requiring that the cantilever be modified or by sliding across a sample.

One general approach lateral force calibration is to determine three conversions. The lateral calibration may be decomposed into an angular sensitivity in rad/V, a torsional spring constant in N–m/rad, and an effective tip height in meters. These three factors may then be combined to produce a complete lateral force calibration. Existing methods are seldom able to independently calculate each of these components, or make numerous assumptions which may not be valid.

The wedge calibration method is a commonly-used standard for comparison [13, 15, 16]; however, it inherently requires a sample change, which is not always suitable for some experimental applications, especially when working in fluid or in AFM setups where changing the sample inherently involves altering the optical alignment. A change in the optical alignment can result in a calibration error of 20% or more [10]. Commonly available commercial wedge gratings based on the silicon (111) planes (TGF11, MikroMasch) are incompatible with colloidal probes with large diameters or with silicon nitride cantilevers in which the angle of the tip is shallower than the angle of the wedge. Moreover, because the wedge method involves sliding on hard surfaces with steep slopes, it has the potential to damage the probe tip; therefore, it is ideally performed at the end of experimentation. This damage can fundamentally alter the probe geometry and affect its calibration, or can even render the calibration data itself unusable [16]. Furthermore, if the calibration is performed at the end of the experiment, the calibration information is not available during the course of the experiment, and no calibration data can be obtained if the cantilever is destroyed during the experiment.

A technique proposed by Wagner *et al.* attempts to achieve a complete calibration of the torsional stiffness and torsional optical lever sensitivity of an AFM cantilever without engaging [16]. However, the method relies on knowledge of the hydrodynamic function of the cantilever [11], which is in general evaluated for a cantilever of rectangular plan view, and which is not strictly valid for most commercial cantilevers. Furthermore, the substantial damping in fluids hinders the hydrodynamic method and its calculation of the hydrodynamic function. Moreover, in order to calibrate lateral force from the torsional spring constant, it is also necessary to determine the tip height. This technique makes no provision for direct *in situ* measurement of the tip height, and instead recommends the use of electron microscopy to measure this parameter. Furthermore, the method is limited to probes for which the hydrodynamic function may be evaluated; the technique is not valid for probes of arbitrary shape such as triangular probes, and may also be inaccurate for probes which feature a large colloid which affects the hydrodynamic function. This method reported an average error of 36% relative to the wedge method, which itself has an uncertainty of at least 12% in addition to the uncertainty sustained due to calibration of the normal spring constant and sensitivity [6]. This uncertainty may be too great for many applications.

A method in which the probe is scanned on a dielectrically suspended sample [13] promises a complete calibration without modification to the tip and without sliding or applying large forces. However, in practice this method is difficult to perform because of complexities involved in setting up and characterizing the oscillation of the sample, which is necessary to determine the spring constant of the assembly. The assembly is quite large and involves powerful magnets, which makes it difficult to install on many commercial AFMs, and it requires a sample exchange, during which the alignment of the detection spot on the cantilever may change. All of these characteristics make this method inconvenient to implement in practice.

Methods such as wedge calibration [8–10] provide a complete calibration from detector voltage to lateral force in Newtons. However, the wedge calibration method requires that the normal spring constant and the normal optical lever sensitivity be calibrated in advance, and it also incorporates these two parameters into its own calibration. Therefore, the uncertainty of the wedge method includes not only its own uncertainty, but also the uncertainty of the methods used to determine the normal spring constant and the normal optical lever sensitivity. The uncertainty of the wedge calibration method may also be greater than anticipated if the tip is unexpectedly damaged or has geometry which interferes with the wedge. Also, the calibration is opaque and offers no convenient means of validation, apart from whether the measured friction coefficients appear to be reasonable.

A method demonstrated by Labuda *et al.* uses an interferometric displacement sensor to measure the torsional spring constant of a cantilever by measuring the thermally driven motion of the cantilever at two or more locations across the width of the cantilever [17]. With a calibrated torsional spring constant, the optical beam detector (OBD) angular sensitivity may also be calculated using the thermal noise spectrum of the OBD lateral signal and the equipartition theorem. However, the authors identified no means of measuring the tip height *in situ*, recommending instead that the tip height be measured using optical microscopy or scanning electron microscopy in order to complete the lateral force sensitivity calibration. Moreover, the interferometric displacement sensor used in that study was an external apparatus which was not fully integrated into the microscope, which complicated its use and prevented the method from being implemented in practice by most users.

We perform lateral force calibration using an AFM (Vero, Oxford Instruments Asylum Research, Santa Barbara, CA) featuring quadrature phase differential interferometry (QPDI) as its primary displacement measurement system [18, 19], while also featuring traditional optical beam detection. With this development, it is now possible to implement a variation of the method of Labuda *et al.* in a general-purpose AFM without any external equipment. Furthermore, we identify a means for measuring the tip height *in situ*, requiring no external measurement methods. Here, we describe a calibration method which is capable of a complete lateral calibration without any cantilever or sample exchanges and which does not require sliding the probe in contact.

## I.    MEASURING TIP HEIGHT

Tip height measurement is accomplished by observing the focus height on both the cantilever and the sample, and by measuring the engage distance [20]. For probes where it is possible to do so, it is ideal to position the detection spot at the same longitudinal position along the cantilever at which the tip is located, as shown by the red beam in Figure 1a. When the detection spot is in this position, no positional corrections to the sensitivity calculation are required apart from the correction to convert from the dynamic to the static spring constant [17]. In practice, one straightforward means to choose the detection spot position is to locate it such that the cantilever displacement observed using the interferometer in a force-distance curve matches the motion of the sample observed by the height sensor. Here, displacement is defined to mean the true displacement of the cantilever as measured by the interferometer. Deflection is the bending of the cantilever measured by OBD, which is in fact an angular measurement from which a displacement is inferred. The force-distance curve approach is applicable when the sample to be measured is significantly more rigid than the probe such that contact stiffness may be neglected, which is true in most lateral force measurements on ceramics and metals. Figure 1 shows the displacement, as measured by the interferometer using its intrinsic calibration to the wavelength of light, as a function of the separation distance between the tip and sample $z_{tip} - z_{sample}$. When this plot is exactly vertical as shown in the red curve, the spot is correctly positioned. The 200 nm spot positioning resolution of this AFM is sufficient to position the spot such that the discrepancy between Z sensor position and the measured interferometric displacement is less than 1% for the models of cantilevers investigated here, all of which feature a tip setback. For probes such as Olympus AC-240 or Oxford Instruments Titan 70 in which the tip is at the end of the lever, or silicon nitride probes such as Nanoworld PNP-DB where the tip is hollow and the detection spot cannot be positioned over the tip, it is necessary to apply a correction factor as will be discussed in Section IV.

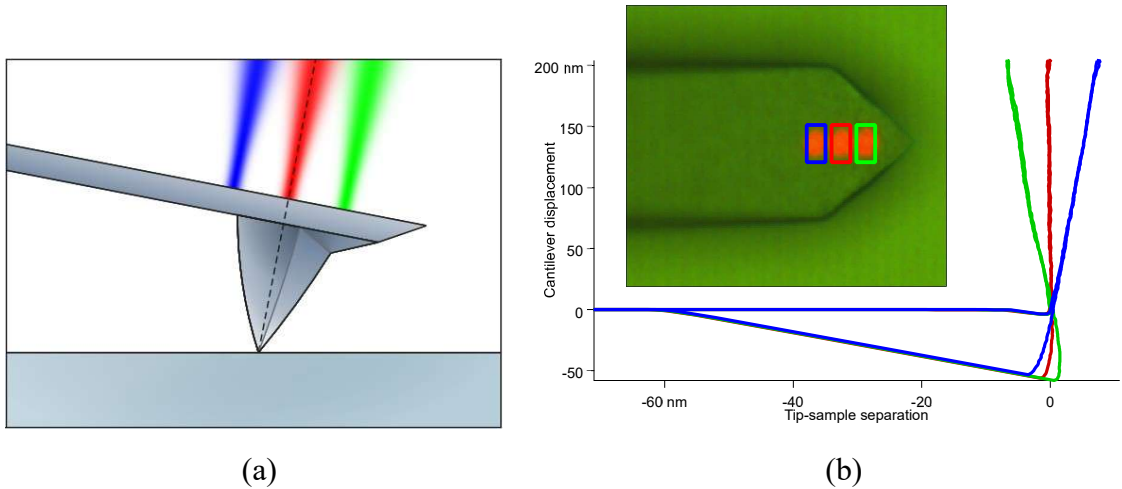

(a)        (b)

FIG. 1: (a) Illustration of the optimum detection spot position (red), with positions which will overestimate (green) or underestimate (blue) the interferometric displacement measurement. (b) plots of displacement versus indentation depth with the detection spot on the left (blue), centered on (red), and to the right (green) of the probe tip. The inset is a composite image showing the spot position for each of the three force curves with the cantilever displacement calibrated using the theoretical value calculated for the interferometer.

The focus height on the tip and on the surface is measured by finding the second moment of the detection spot as viewed by the camera at various focus heights, and takes advantage of the small, focused spot on Vero which is optimized for interferometry. The objective is moved through a range of focus heights such that the spot appears in and out of focus, as shown in 2a. Because the image moment is exquisitely sensitive to even small amounts of light far from the center, this step is performed with ambient illumination off, so the only significant source of light is the detection light source. In addition, several image filtering techniques are applied, including background subtraction, a region of interest, and finally, a seed fill algorithm to eliminate any stray light that is not contiguous with the main spot. Finally, the second moments of the image in both X and Y are taken, as shown in 2b. The location at which the geometric mean of the second moments in X and Y is minimized is determined using a polynomial fit to five data points surrounding the minimum value, and this minimum is taken to be the optimum focus height.

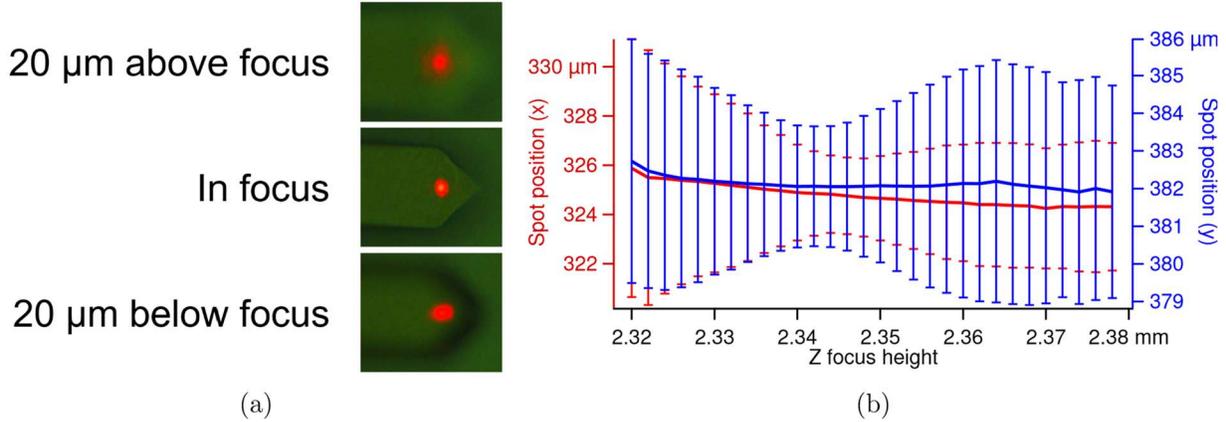

FIG. 2: (a) The detection spot as observed by the optical camera above, in, and below focus. (b) The position (first moment) and variance (second moment) of the detection spot in the field of view as the spot moves through focus.

Because the focus heights described above are ideally performed while the probe is out of contact, it is also necessary to measure the engage distance. If the sample is sufficiently stiff and the detection spot is over the tip, it is sufficient to use the engage distance as measured by the height sensor and to subtract the setpoint in nanometers, as measured by the interferometer. If the sample is soft, it is ideal to use the point at which the probe is in contact with zero force (the corner in Figure 1). If the detection spot cannot be positioned over the tip, then a correction factor accounting for the tilt of the cantilever should be applied.

In order to accurately measure the effective height of the sample, it is necessary to know the neutral axis height rather than the total height of the probe and cantilever; however, this offset is relatively small. The height of the neutral axis is equal to half the cantilever thickness in the case of cantilevers with a rectangular cross-section, or less than half in the case of cantilevers with a trapezoidal cross-section. The uncertainty due to manufacturing variability may be expected to be significantly less than half the thickness. For this reason, we propose using a standard value for the height of the neutral axis based on scanning electron microscopy or the manufacturer's specifications, which may be taken to be half the cantilever thickness for cantilevers with a rectangular cross-section, or 5/12 for cantilevers with a trapezoidal cross-section. The value of 5/12 is chosen to be halfway between the value of 1/2 for a rectangular cross-section and the theoretical value of 1/3 for a triangular cross-section, which can be considered the most extreme case of a trapezoidal cross-section. For the range of probes measured here, the cantilever thickness was less than 20% of the total height of the cantilever and tip. The maximum possible uncertainty due to the assumption made here is only 1/12 of the thickness of the cantilever, which is a few hundred nanometers in most cases and sufficiently smaller than the total tip height to be a negligible source of uncertainty.

In principle, a similar height measurement could be achieved using the interferometric detector itself. In practice, however, the angle of the surface would have to match the angle of the cantilever such that the reflected beam would return at the same angle when reflected off the cantilever or the sample, such that it would interfere with the reference beam and return to the photodetector. Even if this obstacle was overcome, the height of most AFM probes is usually many times the wavelength of light, which is difficult to measure with a single-point QPDI detector. It would, however, be possible in principle to measure the tip height with a white-light interferometer.

## II.   CALIBRATING TORSIONAL STIFFNESS

The torsional stiffness calibration is accomplished using the method of [17], with some modifications to allow for automation. In particular, a thermal spectrum on the OBD lateral signal is collected prior to the interferometric thermal spectra, such as the one shown in Figure 3a. In this case, the first torsional eigenmode is the only visible peak; even in cases where higher-order torsional peaks are visible or normal peaks exist due to crosstalk, the first torsional eigenmode is by far the most prominent. This step allows for unambiguous identification of the first torsional resonance frequency, which is critical because the interferometric detector is sensitive to both normal and torsional modes and features several peaks corresponding to normal modes, as shown in Figure 3b. In addition to being used to identify initial guesses for the fit range, frequency, and quality factor corresponding to the first torsional eigenmode, this OBD thermal spectrum is also used to determine the angular sensitivity as described in Section III.

A running median filter is applied to the power spectral density (PSD) to reduce the effect of scatter on the results, and to eliminate any narrow peaks caused by electrical interference. For convenience, and to eliminate any possibility that the peak finding algorithm will identify acoustic or 1/f noise, we search only for peaks above 10 kHz, which is sufficiently low to include all probes used in this study. The thermal PSD is fit to the following function of frequency *f*:

$$PSD(f) = \sqrt{\frac{\left(\frac{DC\omega_0}{f}\right)^2}{\left(\frac{\omega_0}{f}-\frac{f}{\omega_0}\right)^2+\frac{1}{Q^2}} + WN^2} \qquad (1)$$

where $\omega_0$ is the resonance frequency in Hz, *DC* is the response amplitude extrapolated to a frequency of 0 Hz, and *WN* is the white noise floor, both in $V/\sqrt{Hz}$ or $m/\sqrt{Hz}$ [21]. An example of a QPDI spectrogram used for determining the spring constant is shown in Figure 3c. Initial guesses for the fits to the QPDI thermal spectra are the frequency and quality factor determined from the thermal spectrum in Figure 3a. The initial guess for the white noise is 8 $fm/\sqrt{Hz}$, which is a typical QPDI detector noise floor for the present microscope, and which is valid for almost all cantilevers used in contact-mode imaging, regardless of their lengths. With highly accurate initial values for three of the four fitting parameters in Equation 1, only the thermal DC is a fully unconstrained parameter.

We allow the frequency and quality factor to vary from their initial values at the first step of the thermal spectrum, but not on subsequent steps. This is because the QPDI thermal spectrum has a lower noise floor than the OBD thermal spectrum, and we expect the values from the QPDI thermal spectrum will be more precise than those attained previously from OBD. In practice, the fitted frequency matched with < 1% discrepancy between the detectors in all cases, and the quality factor matched with < 2% discrepancy except in cases where the frequency of the torsional eigenmode was very near to that of a normal eigenmode. Moreover, a good fit of *f* and *Q* is possible at the first and last fit locations because the torsional simple harmonic oscillator (SHO) is usually large compared to the white noise and any neighboring eigenmodes, but these parameters may be difficult to determine at intermediate locations where the SHO may be arbitrarily small. By constraining these parameters based on their value from the initial fit, we prevent the fit from diverging. The thermal DC parameter fluctuates by design; this is the parameter which defines the difference in compliance at different locations. While not strictly necessary, we allow the white noise floor to vary at each location because the white noise floor does, in fact, vary by a few percent depending on the complex contrast of the QPDI detector, and the complex contrast itself varies as the detection spot moves across the width of the cantilever due to imperfections in the cantilever geometry.

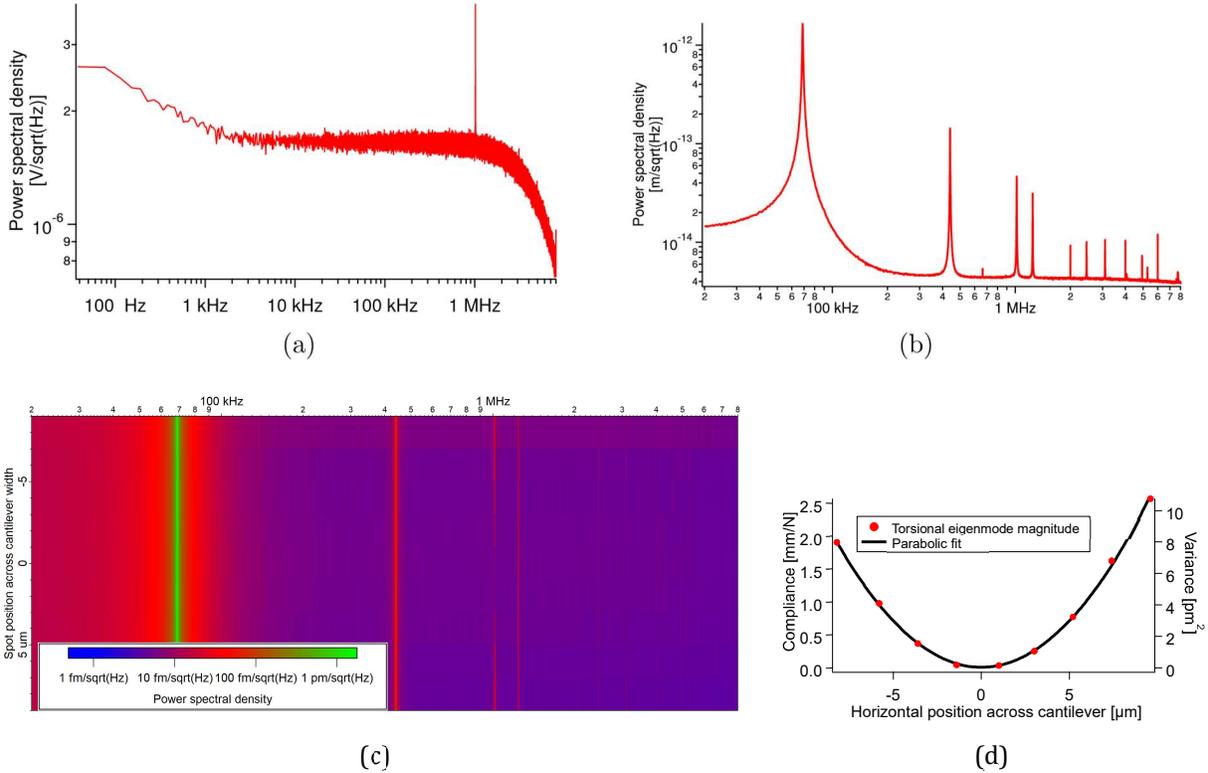

FIG. 3: (a) A thermal spectrum of the OBD lateral signal of an AppNano FORTA probe, used to identify the first torsional resonance frequency and choose a starting value for the quality factor which will be used during the spot position sweep. (b) An example interferometric thermal spectrum, showing the first torsional eigenmode at 1.02 MHz and the second normal eigenmode at 1.24 MHz. (c) A sweep across a range of spot positions. (d) Compliance or variance as a function of spot position across the width of the cantilever.

After fitting each PSD to Equation 1, the compliance $C$ in *m/N* may be calculated for each spot position as

$$C = \frac{\pi \omega_0 Q (DC)^2}{2 k_B T} \quad (2)$$

where $k_b$ is Boltzman's constant and $T$ is the ambient temperature. A second-order polynomial fit to the plot of compliance as a function of position, as shown in Figure 3d, produces the dynamic torsional spring constant $k_{torsional,dynamic}$ of the cantilever in $N-m$. Figure 3d may be compared with Figure 1d of Labuda *et al.* [17]. The static torsional spring constant $k_{torsional,static}$ may be determined as [17]

$$k_{torsional,static} = \frac{8 k_{torsional,dynamic}}{\pi^2} \approx 1.234 k_{torsional,dynamic} \qquad (3)$$

It warrants further mention that the spectrogram in Figure 3c features two neighboring peaks, at 1.02 MHz and 1.24 MHz, corresponding to the second normal eigenmode and the first torsional eigenmode, respectively. It is for this reason that it is beneficial to use the OBD thermal spectrum in Figure 3a to establish the torsional resonance frequency and also establish a narrow fitting window. The larger fitting window that would be required to find the torsional eigenmode would likely also result in a poor fit if it included the adjacent normal eigenmode. Despite the excellent selectivity provided by the current method, it is nevertheless ideal to use probes for which the first torsional eigenmode is not near any other eigenmodes, to provide an additional safeguard against these peaks overlapping due to manufacturing variability across cantilevers, and also to optimize the fit quality.

The calibration method described above could also be implemented using only two points, as described previously [17]. For one of the LFM configurations described in Section V, it is important to determine the offset of the detection spot from the centerline of the cantilever, which is the zero point of Figure 3d. While this is technically possible with the two-point calibration method, it is ideal to perform a calibration at more than two points to achieve greater certainty on the lateral position of the cantilever's neutral axis.

### III. CALIBRATING ANGULAR SENSITIVITY

In order to achieve a complete calibration of the lateral force sensitivity, it is necessary to calculate the sensitivity of the detector to an angular rotation, in *rad/V* . Either the QPDI detector or the OBD detector may be configured to provide lateral sensitivity, as described in the following sections: the lateral channel of the OBD quadrant photodetector may be used; alternatively, the detection spot may be offset to one side of the cantilever and the QPDI detector may be used to measure lateral forces. In this configuration, the QPDI detector is susceptible to crosstalk from normal displacement. However, in some configurations of feedback this crosstalk is not a problem, especially in situations where the trace and retrace images are to be subtracted, or where only the differences in friction forces between two regions of interest or between different lattice sites are of interest. Here, we describe how to calibrate the angular sensitivity for each detector mode.

### A.  OBD lateral detection

A straightforward approach to lateral force measurement using an AFM with both QPDI and OBD detection is to use the QPDI detector to measure and feed back on a constant normal force, while the OBD detector measures lateral displacement. The sensitivity of the OBD detector to lateral displacement may be calculated using the equipartition theorem, as described previously [17]:

$$\frac{1}{2} k_{torsional} (Sensitivity_{torsional})^2 <V_{lat}^2> = \frac{1}{2} k_B T \qquad (4)$$

where $k_{torsional}$ is the torsional spring constant in $N-m/rad$ determined previously; $Sensitivity_{torsional}$ is the angular inverse optical lever sensitivity in $rad/V$, the quantity of interest; and $\langle V^2 \rangle$ is the variance of voltage of the lateral signal near the torsional resonance frequency, as measured by the OBD detector.

A summary of calibration parameters for several probes used in this study is shown in Table I. The angular sensitivity is primarily a function of the optical and electrical design of the microscope and does not change appreciably across a wide range of cantilevers. What variability does occur is primarily due to the reflectivity of the cantilever. Because the tip height and torsional spring constant are properties of the cantilever only, these parameters could be measured in advance on an interferometric AFM. This advance measurement would allow a complete calibration of lateral sensitivity on another microscope, for example an entirely OBD microscope, by finding only the angular sensitivity. Furthermore, separating these three parameters allows for more effective evaluation of the suitability of a cantilever for a particular experiment, for example when considering the criteria laid out for lattice resolution described by Yang *et al.* [22].

TABLE I: A summary of calibration parameters for several probes used in this study. The tip height is measured from the neutral axis of the cantilever. All of the probes used here feature a tip setback, and the spot was positioned over the tip for calibration as shown in Figure 1.

| Probe | Tip height (μm) | Torsional spring constant (nN-m) | Angular inverse sensitivity (mrad/V) | Complete lateral sensitivity (μN/V) |
|---|---|---|---|---|
| FORTA | 13.9 | 29.1 | 2.05 | 3.47 |
| qp-BioAC (long) | 6.7 | 0.250 | 1.68 | 0.0506 |
| qp-BioAC (medium) | 6.6 | 0.299 | 1.60 | 0.0591 |
| qp-BioAC (short) | 6.7 | 0.394 | 1.66 | 0.0807 |
| SICONG | 16.1 | 35.3 | 2.31 | 4.10 |
| ContPt | 15.7 | 27.0 | 1.91 | 2.66 |

### B. QPDI offset spot lateral detection

An alternative lateral force detection option is to position the detection spot offset to one side of the cantilever, and use the QPDI detector to measure the twisting of the cantilever [20]. In this configuration, the OBD normal deflection signal may be used for measuring and feeding back on normal force, or feedback may be turned off entirely. Note that to first order, both the OBD normal and lateral signals are insensitive to the spot position transverse to the cantilever. The QPDI detection signal may be converted to an angle of twist $\theta_{twist}$ as

$$\theta_{twist} = \tan^{-1} \frac{displacement}{d_{offset}} \approx \frac{displacement}{d_{offset}} \quad (5)$$

where *displacement* is the displacement measured by the interferometer and $d_{offset}$ is the horizontal offset to the detection spot from the centerline of the cantilever. It is furthermore possible in this detection mode to use the OBD lateral detection as an additional measurement of the angle of twist, such that the two detectors acquire data simultaneously and their results may be compared.

It is worth noting that this calibration method assumes that the probe tip is centered across the width of the cantilever, which was not true for some of the cantilevers considered in this study due to manufacturing defects.

## IV. EXTRAPOLATION FOR VISIBLE APEX CANTILEVERS

When the detection spot cannot be placed over the tip, for example on visible apex probes or silicon nitride probes, it is necessary to extrapolate the calibration parameters to the tip location. Figure 4 shows how each of the three components of the lateral force calibration vary with spot position. Each of these parameters may be independently extrapolated to the tip based on their expected functional form. The dynamic torsional spring constant (Fig. 4a) compares favorably with Figure 3 of [17], which shows a trend in which the dynamics of the cantilever become nearly flat toward the end, due to the lack of a significant inertial load beyond the detection spot, which is also the point at which the dynamics are being probed. Meanwhile, the angular sensitivity (Fig. 4b) does not depend on the spot position, which is to be expected on any cantilever with a uniform coating. Indeed, the value of the angular sensitivity was broadly consistent across all cantilevers investigated in this study on the same microscope as shown in Table I, although it can be expected to vary on cantilevers of differing reflectivity. The reflectivity of the cantilever may be estimated using the OBD sum, also shown in Figure 4b.

The tip height (Fig. 4c) shows a linear trend with a slope approximately equal to $\sin(11°)$, the tilt of the cantilever. Interestingly, no effort was made to correct the sensitivity of the interferometer to account for the different spot position; with a sufficiently small setpoint this effect has a negligible impact on the measured tip height in comparison with the tilt of the cantilever. Therefore, if the spot position and cantilever tip position are both known, along with the length of the cantilever, it is not challenging to extrapolate each of these parameters to the position of the tip. The lateral signal in volts should also be extrapolated; this extrapolation should be proportional to position along the cantilever for rectangular cantilevers, as long as the scan frequencies are sufficiently below the in-contact torsional resonance of the cantilever such that the response of the cantilever may be assumed to be that of a quasistatic spring. The variation of the overall sensitivity (Fig. 4c is nonmonotonic because it is dependent upon both the tip height and the measured spring constant, which both vary monotonically with spot position.

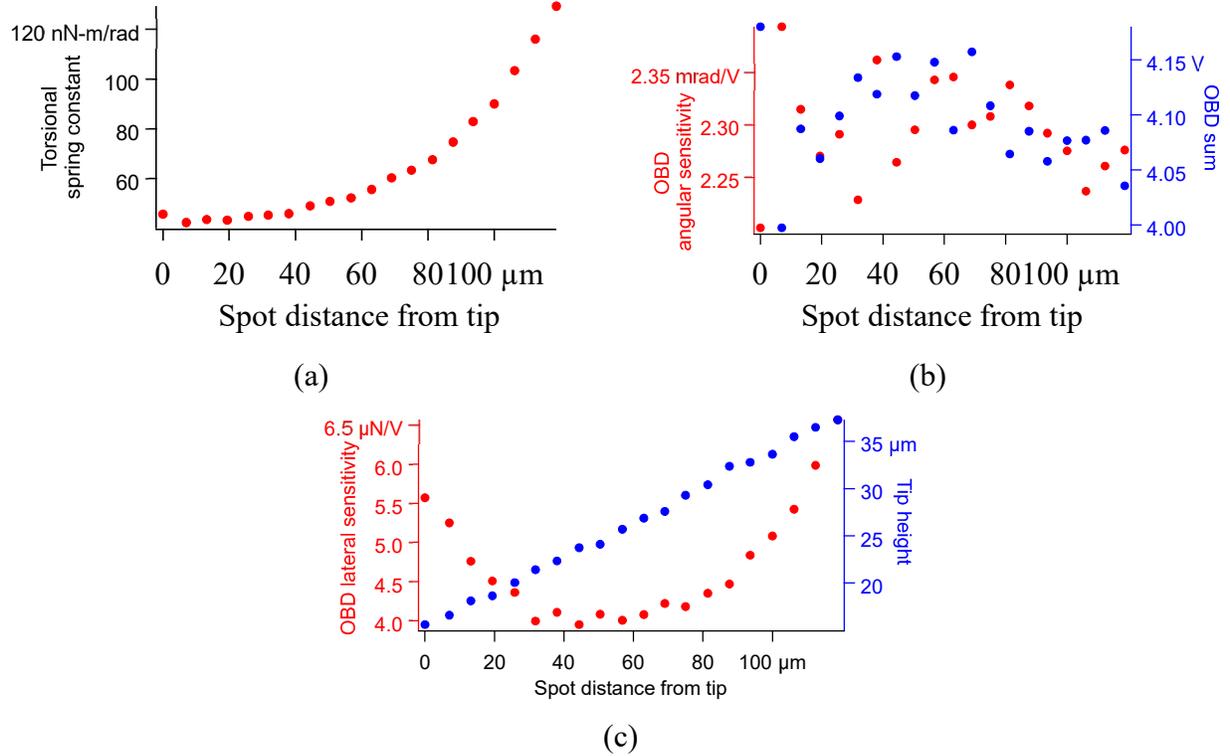

FIG. 4: Effect of spot position on (a) the measured dynamic torsional spring constant for a FORTA probe, (b) the angular sensitivity, and (c) the measured tip height and complete lateral force sensitivity.

## V. TWO ALTERNATIVES FOR LATERAL FORCE MEASUREMENT WITH QPDI

As described in Section III, the QPDI signal may be used as a measurement of either the normal force or the lateral force. These were not possible using the laser Doppler vibrometer employed previously [17], because the latter could not stably measure a constant offset in the signal. The noise performance of the QPDI detector is superior to the noise performance of the OBD detector both in and out of contact, as shown in Figure 5. The normal eigenmode appears in the QPDI lateral signal when out of contact (Fig. 5a); however, in contact this oscillation is essentially nullified by positioning the detection spot near the tip. Moreover, the in-contact normal eigenmode shifts to a much higher frequency, which can be filtered so that it does not affect measured data. The noise in the QPDI lateral signal in contact (Fig. 5b) corresponds to coupling from mechanical oscillations in the scanner and sample. These oscillations may possibly be vertical, and appear in the QPDI signal because it is sensitive to both vertical and horizontal fluctuations in this detection scheme.

Because this noise is mechanical, it does not appear when out of contact, and also does not appear in the OBD lateral signal due to its higher noise floor and because the OBD lateral signal is not sensitive to vertical oscillations to first order. The superior noise floor of the QPDI detector indicates that it will outperform the OBD detector, especially at high speeds. We note, however, that if the OBD photodetector is used for rapid feedback on OBD deflection, then the noise of the OBD photodetector is coupled to the QPDI detector and can worsen the performance of the latter. Therefore, we recommend that the offset spot detection configuration be used primarily in situations where height feedback is not necessary and may be slowed down or turned off, such as for rapid scanning of small areas, or in two-pass modes in which feedback is not enabled during the lateral force scan.

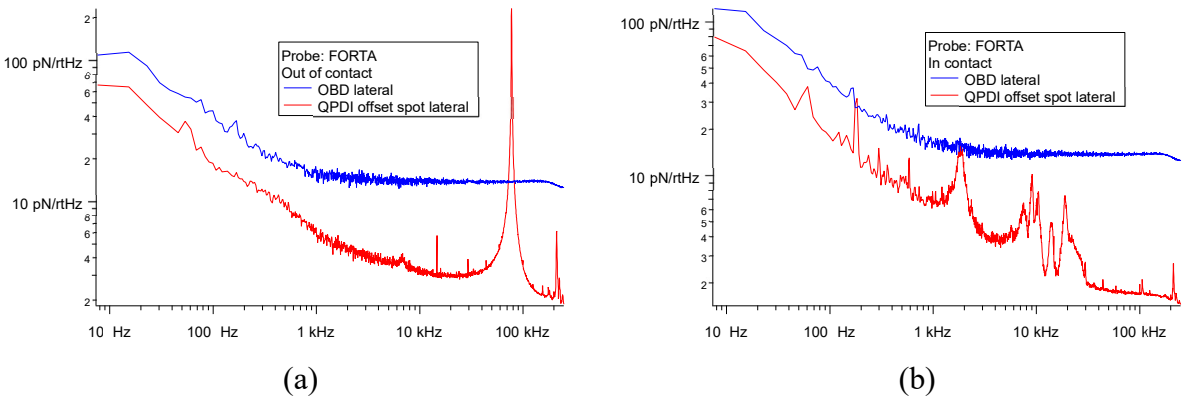

FIG. 5: A direct comparison of the noise performance of OBD lateral and QPDI offset spot lateral for a spot offset of 10 um (a) out of contact, and (b) in contact with feedback on QPDI and a low integral gain of 0.01.

The two methods described here have different advantages for different experiments. For example, the QPDI offset spot method is capable of arbitrarily high sensitivity when used with a cantilever of sufficient width. However, the QPDI approach also introduces substantial crosstalk between the normal displacement of the cantilever and the QPDI lateral signal. The OBD lateral detection method does not exhibit this crosstalk and is capable of functioning on a narrow cantilever such as a NuNano Scout 70; however, the calibration method described here is of limited usefulness in this case because the lateral motion of the spot is limited. Both detection methods are capable of achieving atomic resolution on highly ordered pyrolytic graphite, as shown in Figure 6, and in particular the interferometric detector produces a slightly improved noise floor as evidenced by the decreased amount of serrations shown in Figure 6c. A thorough discussion of the benefits of each detection method is outside the scope of the current article; here, we concern ourselves primarily with how to apply the results of the calibration to each of these two methods.

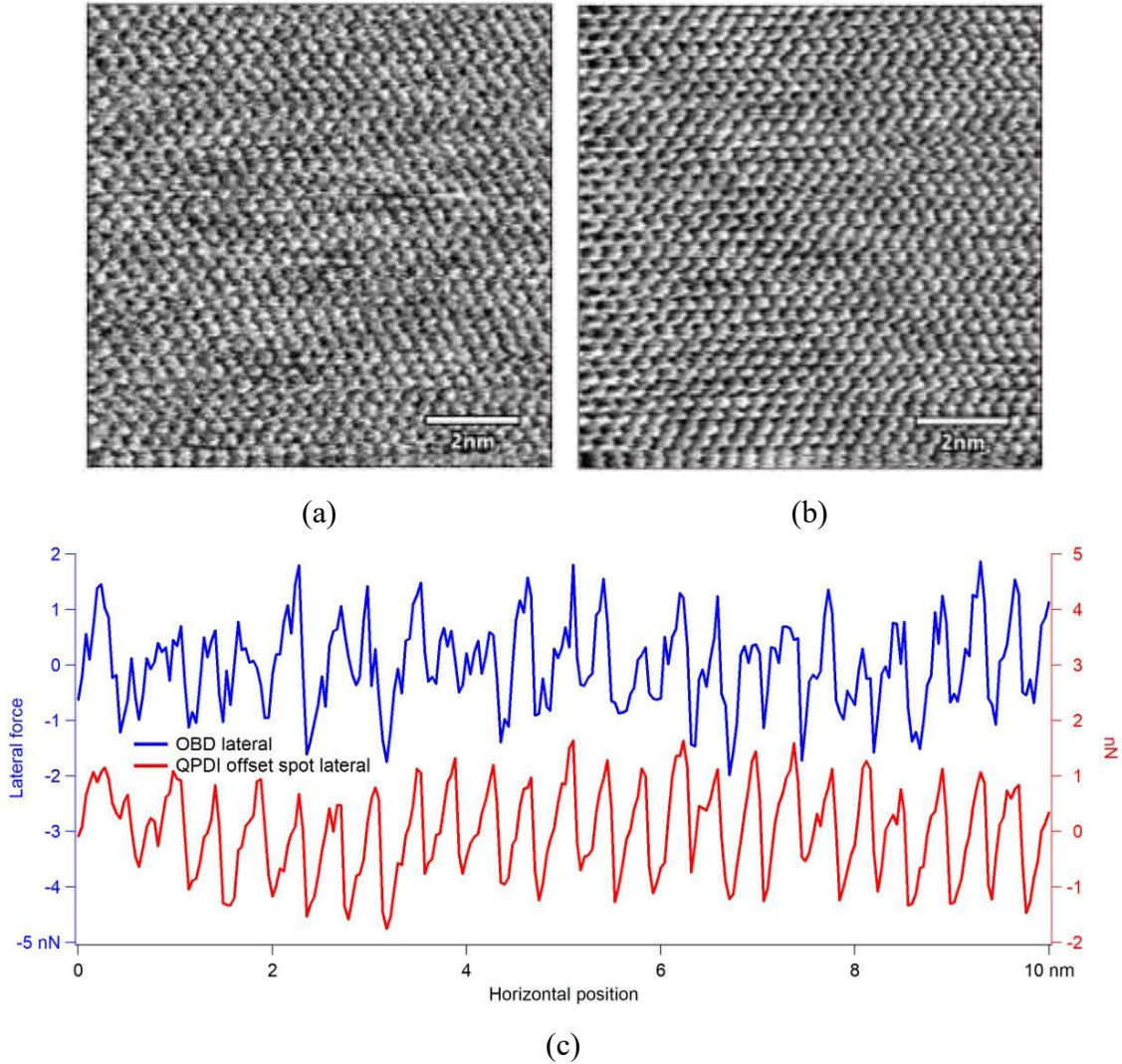

FIG. 6: Atomic lattice resolution on tungsten diselenide in contact mode using (a) optical lever detection and (b) interferometric detection and an offset spot. (c) Line traces from each image, showing the same features but with slightly improved noise performance on the interferometric trace. The two images were captured concurrently.

## VI. VALIDATION AND UNCERTAINTY ANALYSIS

The accuracy of the tip height measurement described in Section I is confirmed by direct comparison with scanning electron microscopy, as shown in Figure 7. Here, both the electron microscope measurement and the AFM measurement were used to determine the height of the tip from the top surface of the cantilever, rather than from the neutral axis which is the quantity of interest, because this former measurement is the only dimension which may be measured directly using either technique. Across nine probes of three models, the two measurements always matched within 1.4 µm and in almost all cases to within 600 nm. Some of this uncertainty may be attributed to a combination of the encoder resolution of the AFM, random fluctuations in the data, and human error in performing the SEM measurement. We note that the SEM measurement is consistently larger than the AFM measurement by a few percent; this may correspond to a small calibration error in either of the instruments used in the study.

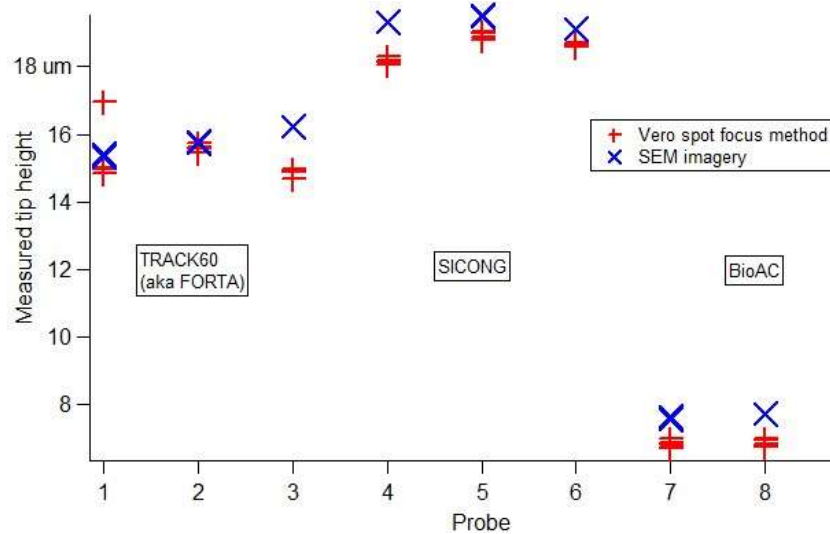

FIG. 7: Comparison of tip heights for three different models of AFM cantilevers as measured by scanning electron microscopy, and using the method based on spot focus height described here. Each probe was measured twice by electron microscopy and five times using the spot focus height method.

Across the selection of probes listed in Table I, the interferometric calibration method described here produced a lateral calibration which ranged from 4% greater to 23% less than the result produced by the wedge method [8]. In our experiments, we observed that the wedge method itself can have an uncertainty of 50% or more, depending on the range of forces applied during the calibration experiments. Through repeated experimentation at a variety of load ranges, we found that most lateral sensitivity values measured by the wedge method were clustered within ±10% uncertainty with only a few outliers, as shown in the Supporting Information.

In those instances where the lateral sensitivity determined by the interferometric method differed from the wedge method by more than 20%, it was not within the range of values which the wedge method produced, even including the outliers. This finding means that it is unlikely that the discrepancies we observe may be attributed solely to the intrinsic uncertainty of the wedge method resulting from the range of loads that are applied.

Of the five models of cantilever for which data is presented here and several others that are not shown, in most instances where the discrepancy between the interferometric calibration and the wedge calibration was the greatest, we observed that the first torsional eigenmode frequency was very close to one of the normal eigenmodes. We hypothesize that the discrepancy in calibration values may be due to poor fit quality or mixing between the normal and torsional eigenmodes. Only one instance (Mikromasch DPER-XSC11) was identified in which the calibration produced by the wedge method differed substantially from that of the interferometric method, and for which the first torsional eigenmode was not close to one of the normal eigenmodes. On that probe model, the cantilever is only approximately 18 μm wide at the position of the tip, which limits the range of motion of the detection spot to about 15 μm and produces substantial uncertainty in the measured spring constant.

We additionally attempted to compare with the dielectric levitation calibration method [13]; however, we found this method impractical in our experimentation. In particular, it was not feasible to identify an orientation of the assembly in which the levitated sample oscillated stably in one direction without also rotating or oscillating in the perpendicular direction. Moreover, a single consistent resonance frequency could not be established. We chose to disregard the dielectric levitation calibration method for comparison purposes.

## VII. BENEFITS

The QPDI detection used here is stable at low frequencies, which allows for the measurement of nominally constant displacements and torsional bending, as well as slowly varying signals. By contrast, the laser Doppler vibrometer used by Labuda *et al.* does not produce a stable output at low frequencies and cannot clearly resolve thermal noise spectra below 10 kHz. The QPDI detector also has a lower noise floor and is therefore capable of achieving lower uncertainty than observed previously. Furthermore, since the QPDI and OBD detectors on Vero make use of a single detection beam, it is not necessary to manually align the beams so that they overlap as was necessary in the previous apparatus. It is furthermore possible to use the interferometric detector to exactly position the beam directly over the tip, which eliminates uncertainty due to spot position. On cantilevers where the spot cannot be placed directly over the tip, it is still possible to extrapolate to determine a conversion factor to correct the sensitivity [17, 23]. The method described here is fully automated, requiring only the height of the neutral axis of the cantilever as an input, which may come from a database and which has a small effect on the uncertainty of the final calibration. In addition to its obvious use for lateral force microscopy, a variation of the technique described here could also be useful for calibration for torsional resonance and vector measurement modes [20].

The wedge method [8] incorporates all of the uncertainty in the measurement of the normal sensitivity and normal spring constant, which can be high for OBD, especially for the long cantilevers commonly used in LFM measurements. The wedge method also does not provide any useful metrics apart from the measured coefficients of friction which may be used to validate the quality of the result. These coefficients are dependent on the tip condition and cannot be easily measured by any other means, and furthermore are not always accurate indicators of the quality of the calibration as outlined in the Supporting Information. Conversely, the present method provides several individual, meaningful parameters that can be compared with external measurements. For example, the angular sensitivity may be compared with the result of a colloid pressed laterally against a wall [14]. Having established the angular sensitivity, the spring constant may be compared directly with that calculated from an OBD lateral thermal spectrum. Finally, the tip height may be measured directly with an SEM as was done here.

During the development of this calibration method, we performed comparisons with both the wedge method [8] and the dielectric spring method [13], and identified numerous challenges associated with both of those methods, of which only some are acknowledged in the existing literature. Principally, both of these calibration methods require a sample change, which may affect the spot alignment on the tip and therefore the calibration. This may complicate the calibration workflow, especially on AFMs configured for small samples.

As mentioned previously, the wedge calibration method is not compatible with colloidal probes or probes with large cone angles. If the point of contact between the probe tip and the wedge is not near the end of the tip, then the wedge calibration will be in error. This condition also raises the possibility that a calibration might fail on a probe with a sufficiently sharp cone angle if the probe is damaged during the experiment. Moreover, the wedge calibration method is sensitive to surface contamination on the wedge, as well as feedback settings that are used. It is also imperative that the wedge position and the vertical range of motion of the AFM are sufficient to allow the desired maximum force to be attained, especially with cantilevers with low spring constants in which substantial Z travel is required to apply large loads.

The dielectric levitation method of calibration presents even further obstacles, as the powerful magnets may attract to other components of the AFM or attempt to pull the configuration into particular orientations. This constraint makes it challenging to position the tip over the center of the graphene flake or to implement Li *et al.*'s recommendation to orient the system in the direction in which the oscillation is the most stable. Indeed, we could not find an orientation in which the sample oscillated in a purely linear direction. Furthermore, the measured resonance frequency of the graphene flake is challenging to measure accurately. During the present experiments, a variation in frequency of 30% was observed depending on whether the system was actuated by using the piezoelectric actuator in the scanner or by tapping on the enclosure. This discrepancy results in over 50% uncertainty in the spring constant, which would translate to any lateral force calibration performed using this spring constant. We speculate that the tap on the enclosure may also cause the AFM to move on its vibration stabilization stage, which may contribute to the observed frequency response. We believe that actuating the sample using the piezoelectric actuator provides a more reliable perturbation; however, on an AFM equipped with a tip scanner rather than a sample scanner, it would be impossible to actuate the sample in this way *in situ*. Since the resonance frequency may be different if graphene is modified, cleaved, damaged, or even rotated, it is advisable to perform this measurement of mass, frequency, and spring constant every time the method is implemented. Therefore, the logistics of performing an up-to-date *ex situ* measurement compound the complexity of this method. We also note that the graphene flake may drag on the magnets, or the probe may be positioned off-center such that the graphene is rotating in addition to translating, which adds further uncertainty to this measurement.

A problem with conventional lateral force measurements described previously [1], in which the normal force signal fluctuates due to longitudinal bending, may be decoupled using the dual-detector technique described here. The QPDI detector may be used to unambiguously measure tip height, while the OBD photodetector may be used to measure the lateral signal and potentially also the longitudinal bending.

We note that the method of measuring the lateral force using the QPDI detector with an offset spot assumes that the probe tip is positioned on the centerline of the cantilever. If the spot is offset due to manufacturing errors, then the moment arm of the interferometric measurement will be offset. We recommend imaging probes under an optical microscope to confirm that the tip appears symmetric to avoid this problem; otherwise, the calibration could be manually corrected to account for the tip offset.

The basic method described here is applicable in liquid as well as in air, provided that the interferometric sensitivity is corrected for the refractive index of the medium. In practice, the broadening and shifting of the resonance peaks due to damping can cause a loss of fit quality as described in Section II. The efficacy of the method in liquid depends on the probe, the frequencies of neighboring eigenmodes, and the size of the torsional eigenmode as observed in the OBD lateral thermal spectrum. For this reason, a fully automated approach is not recommended in liquid. Instead, we recommend that the tip height and torsional spring constant are first calibrated in air, and then the tip is submerged in liquid for calculating the angular sensitivity. Alternatively, the fully automated approach may be used, provided that the user validates the fit quality manually.

## VIII. CONCLUSION

We describe a new calibration method for lateral force microscopy making use of a combined interferometric and optical beam detector, which is based on a previous technique [17] which performs a partial calibration and makes use of additional capabilities of the AFM to additionally measure the tip height, an important factor which was not fully addressed in the prior work. Keeping in mind the practical complexities associated with both the wedge method [8] and the dielectric suspension method [13], the present calibration method may be performed in a similar amount of time and with comparable accuracy. In contrast with these prior methods, however, it may be performed without sample exchanges and with minimal user intervention.

In addition, two methods are described for lateral force operation. One method operates similarly to conventional lateral force microscopy making use of the quadrant photodetector for the lateral signal, but uses the interferometric signal for the normal signal. This method has a benefit over conventional detection in that if the detection spot is positioned accurately above the tip, then crosstalk from torsional bending into the normal signal is eliminated.

A second method uses the interferometric detector and an offset spot to measure the lateral signal, which provides better noise performance especially at higher scan rates, with a caveat that normal displacement of the cantilever couples into the lateral signal as well. This problem can be remedied by using both the optical beam detector and the interferometric detector simultaneously, such that the optical beam detector measures the absolute magnitude of the lateral signal, whereas the interferometric detector provides better noise performance at high speeds, for example for lattice resolution. The calibration method described here can be applied to both of the described detection methods.

The opportunity for an automated calibration method along with new options for detection make a combined interferometric and optical beam AFM an attractive option for lateral force microscopy. We anticipate that this development will be beneficial especially for experiments where calibrated, high-speed data capture is beneficial, for example for atomistic wear investigations and rate-and-state models of friction. It will also be useful in other lateral force microscopy studies in which the currently-existing calibration techniques are precluded by experimental considerations.

## IX. ACKNOWLEDGEMENTS

The authors gratefully acknowledge support from Chris Santeufemio for imaging the probes in Figure 7, and Dara Walters and Ted Limpoco for useful discussions.

**REFERENCES**

[1] S. Fujisawa, Y. Sugawara, S. Ito, S. Mishima, T. Okada, and S. Morita, The two-dimensional stick-slip phenomenon with atomic resolution, Nanotechnology **4**, 138 (1993).

[2] H. Hölscher, D. Ebeling, and U. D. Schwarz, Friction at atomic-scale surface steps: Experiment and theory, Physical Review Letters **101**, 246105 (2008).

[3] C. Lee, Q. Li, W. Kalb, X.-Z. Liu, H. Berger, R. W. Carpick, and J. Hone, Frictional characteristics of atomically thin sheets, Science **328**, 2562 (2010).

[4]  J. B. McClimon, Z. Li, K. Baral, D. Goldsby, I. Szlufarska, and R. W. Carpick, The effects of humidity on the velocity-dependence and frictional ageing of nanoscale silica contacts, Tribology Letters **72** (2024).

[5]  H. I. Kim, T. Koini, T. R. Lee, and S. S. Perry, Systematic studies of the frictional properties of fluorinated monolayers with atomic force microscopy: Comparison of $CF_3$-and $CH_3$-terminated films, Langmuir **13**, 7192 (1997).

[6]  M. Munz, Force calibration in lateral force microscopy: a review of the experimental methods, Journal of Physics D: Applied Physics **43**, 063001 (2010).

[7]  M. L. Palacio and B. Bhushan, Normal and lateral force calibration techniques for afm cantilevers, Critical Reviews in Solid State and Materials Sciences **35**, 73 (2010).

[8]  D. F. Ogletree, R. W. Carpick, and M. Salmeron, Calibration of frictional forces in atomic force microscopy, Review of Scientific Instruments **67**, 3298 (1996).

[9]  M. Varenberg, I. Etsion, and G. Halperin, An improved wedge calibration method for lateral force in atomic force microscopy, Review of Scientific Instruments **74**, 3362 (2003).

[10] H. S. Khare and D. L. Burris, The extended wedge method: Atomic force microscope friction calibration for improved tolerance to instrument misalignments, tip offset, and blunt probes, Review of Scientific Instruments **84**, 055108 (2013).

[11] C. P. Green, H. Lioe, J. P. Cleveland, R. Proksch, P. Mulvaney, and J. E. Sader, Normal and torsional spring constants of atomic force microscope cantilevers, Review of Scientific Instruments **75**, 1988 (2004).

[12] C. Cafolla, A. F. Payam, and K. Voïtchovsky, A non-destructive method to calibrate the torsional spring constant of atomic force microscope cantilevers in viscous environments, Journal of Applied Physics **124**, 1 (2018).

[13] Q. Li, K.-S. Kim, and A. Rydberg, Lateral force calibration of an atomic force microscope with a diamagnetic levitation spring system, Review of Scientific Instruments **77**, 065105 (2006).

[14] K. H. Chung, J. R. Pratt, and M. G. Reitsma, Lateral force calibration: Accurate procedures for colloidal probe friction measurements in atomic force microscopy, Langmuir **26**, 1386 (2010).

[15] R. J. Cannara, M. Eglin, and R. W. Carpick, Lateral force calibration in atomic force microscopy: A new lateral force calibration method and general guidelines for optimization, Review of Scientific Instruments **77**, 053701 (2006).

[16] K. Wagner, P. Cheng, and D. Vezenov, Noncontact method for calibration of lateral forces in scanning force microscopy, Langmuir **27**, 4635 (2011).

[17] A. Labuda, C. Cao, T. Walsh, J. Meinhold, R. Proksch, Y. Sun, and T. Filleter, Static and dynamic calibration of torsional spring constants of cantilevers, Review of Scientific Instruments **89**, 093701 (2018).

[18] L. Bellon, S. Ciliberto, H. Boubaker, and L. Guyon, Differential interferometry with a complex contrast, Optics Communications **207**, 49 (2002).

[19] P. Paolino, F. A. Aguilar Sandoval, and L. Bellon, Quadrature phase interferometer for high resolution force spectroscopy, Review of Scientific Instruments **84**, 095001 (2013), 1306.0871.

[20] R. Proksch and R. Wagner, Accurate 3d nanoscale electromechanical imaging with a metrological atomic force microscope, Small Methods (2025), in press.

[21] A. P. French, *Vibrations and Waves* (CRC Press, New York, 1971).

[22] Y. Yang, K. Xu, L. N. Holtzman, K. Yang, K. Watanabe, T. Taniguchi, J. Hone, K. Barmak, and M. R. Rosenberger, Atomic defect quantification by lateral force microscopy, ACS Nano **18**, 6887 (2024).

[23] A. Labuda, M. Kocun, M. Lysy, T. Walsh, J. Meinhold, T. Proksch, W. Meinhold, C. Anderson, and R. Proksch, Calibration of higher eigenmodes of cantilevers, Review of Scientific Instruments **87**, 1 (2016).